\documentclass[conference]{IEEEtran}
\IEEEoverridecommandlockouts

\usepackage{cite}
\usepackage{amsmath,amssymb,amsfonts}
\usepackage{graphicx}
\usepackage{textcomp}   
\usepackage{xcolor}
\usepackage{subcaption}
\usepackage{booktabs}
\usepackage{gensymb}

\usepackage[hidelinks]{hyperref}

\begin{document}

\title{Development of ML model for triboelectric nanogenerator based sign language detection system  }

\author{
\IEEEauthorblockN{Meshv Patel\IEEEauthorrefmark{3}\thanks{Note: Meshv Patel and Bikash Baro contributed equally to this work.}, Bikash Baro\IEEEauthorrefmark{2}, Sayan Bayan\IEEEauthorrefmark{2}{*}, Mohendra Roy\IEEEauthorrefmark{3}{*}}
\IEEEauthorblockA{\IEEEauthorrefmark{3}Department of ICT, Pandit Deendayal Energy University, Gandhinagar, India}
\IEEEauthorblockA{\IEEEauthorrefmark{2}Department of Physics, Rajiv Gandhi University, Rono Hills, Itanagar, Arunachal Pradesh \\
*Corresponding Authors: mohendra.roy@ieee.org ;  sayan.bayan@gmail.com }
}

\maketitle

\begin{abstract}
Sign language recognition is important since there are no limits to the use of translators between deaf and hearing communities, and vision-based approaches that are already in place suffer from sensitivity to objects that block the view, difficult computational tasks, and physical considerations. This work presents a comprehensive comparison between ML and deep learning models for a custom developed triboelectric
nanogenerator-based sensor-powered glove for sign language recognition, utilizing multivariate time-series data from five flex sensors. This work systematically benchmarks traditional ML algorithms, feedforward neural networks, LSTM-based temporal models, and a multi-sensor MFCC CNN-LSTM architecture across 11 sign classes (digits 1-5 and letters A-F). The proposed MFCC CNN-LSTM architecture, in which each sensor's frequency-domain features are processed by independent convolutional branches before fusion, achieves 93.33\% accuracy and 95.56\% precision, representing a 23 percentage point improvement over the best traditional ML algorithm (Random Forest: 70.38\%). Ablation studies reveal that 50-timestep windows represent a reasonable tradeoff between temporal context and training data volume with 84.13\% accuracy compared to 58.06\% with 100-timestep windows; MFCC feature extraction maps temporal variations to execution-speed-invariant spectral representations; and data augmentation methods (time warping, magnitude scaling, temporal shifting, noise injection) are essential for generalization. Results demonstrate that frequency-domain feature representations combined with parallel multi-sensor processing architectures offer considerable enhancement over classical algorithms and time-domain deep learning algorithms for wearable sensor-based gesture recognition with practical implications for assistive technology development.
\end{abstract}

\begin{IEEEkeywords}
triboelectric
nanogenerator, Sign Language Recognition,, Machine Learning, Deep Learning, Wearable Sensors, Gesture Recognition, Multi-Sensor Processing
\end{IEEEkeywords}

\section{Introduction}
\label{sec:introduction}

Sign language recognition systems can help bridge the communication gap between the deaf and hearing communities. Traditional vision-based algorithms are highly limited by occlusion, accentuation of variations, sensitivity to camera quality, background clutter, focus problems, and computational complexity~\cite{bragg2019sign,rastgoo2021sign}. These vision-based solutions are not suitable for real-world situations, as they are anticipated to be used in a controlled environment with brightly illuminated, high-resolution cameras and a clear view. Recognition can also be severely harmed by changes in camera angles, camera distance, and environmental conditions. A data glove-based system that allows for the direct measurement of finger and hand movements through wearable devices is one of these potentially beneficial alternatives~\cite{dipietro2008survey}. It can be applied with a significant degree of success even without the need to use visual conditions, changes in luminance, and camera characteristics.

In this study, ML and deep learning methods are examined to recognize sign language movements by using multisensor data. In addition, various neural network topologies are examined as well as specialist temporal models to facilitate the recognition of the movements. Time-series classification problems have shown promising results with recent advances in deep learning ~\cite{fawaz2019deep}. The study includes: (1) a detailed comparison between deep learning and classical methods; (2) an evaluation of feature extraction techniques, such as MFCC representations, with data augmentation strategies ~\cite{iwana2021empirical}; (3) a systematic evaluation of window size and normalization effects; and (4) development of multi-sensor CNN-LSTM algorithms with parallel processing.

\section{Related Work}
\label{sec:related}

It has been observed that vision-based methods and sensor-based methods are the two main approaches to sign language recognition. Vision systems \cite{koller2019weakly} are camera-based and therefore rely on visual input to detect gestures; however, they are computationally challenging, sensitive to lighting conditions, and prone to occlusion. Data-glove systems measure the kinematics of the hands and fingers using flex sensors and inertial measurement units \cite{dipietro2008survey}. In other studies, classical ML techniques such as Support Vector Machines and Hidden Markov Models \cite{starner1998real} have been applied using manually designed features. However, these techniques require extensive feature engineering and are unable to accurately represent complex temporal relationships.

Recent developments in deep learning have enhanced time-series classification \cite{fawaz2019deep}. LSTM networks are said to offer better sequence gesture recognition capabilities than long-term time dependent relationship models \cite{graves2013speech,ordoñez2016deep}. Hybrid CNN-LSTM models are created by combining time modelling and spatial feature extraction of wearable sensor data \cite{ordoñez2016deep}. The frequency-domain representation, which has not yet been shown to be sensitive to changing timing implementation of the activity \cite{ravi2016deep}, was found to be popular in sensor-based activity recognition, and the MFCC features, which were initially designed to aid with speech recognition \cite{mfcc1}, have proven helpful in sensor-based activity recognition. Temporal warping data augmentation and noise injection have proved essential for enhancing model generalisation when utilising small quantities of training data \cite{iwana2021empirical,um2017data}. Despite these advances, comprehensive systematic comparisons across multiple architectures and feature representations for data glove-based sign language recognition remain limited. This work addresses this gap by custom-fabricated nanogenerator-based glove and by doing extensive evaluation across traditional ML, feedforward networks, LSTM-based models, and specialized MFCC CNN-LSTM architectures, with systematic ablation studies on window size, normalization, and augmentation strategies.

\section{Methodology}
\label{sec:methodology}

\subsection{Synthesis of ZnO}
The straightforward chemical method was used to create the ZnO samples. First, a seed solution of sodium hydroxide (NaOH) and zinc acetate dihydrate (ZAD, Zn(CH3CO2)2.2H2O) was made with methanol in a 3:1 molar ratio after being stirred for an hour at 50 OC. The naked cotton cloth was then dried for an hour at about 80 degrees Celsius after being submerged in the prepared seed solution for ten minutes. 
The dried cotton substrate with ZnO seed layer was then submerged in a solution of equimolar zinc nitrate hexahydrate (ZNH, Zn(NO3)2.6 H2O) and hexamethylenetetramine (HMT, (CH2)6N4) in 40 mL of deionized water, and it was heated to 90 °C for two hours in a closed environment to aid in the ZnO growth process. The final product in the form of white film deposited on the flexible cotton substate was washed with deionized water and dried at 60\degree C for 30 min 

\subsection{Fabrication of STENG or sensor}
A single electrode triboelectric nanogenerator (STENG) has been fabricated using ZnO nanorods. In this design, aluminum foil serves as one of the counter layers, which is attached with adhesive polypropylene tape. The ZnO is grown on a cotton substrate, and the aluminum foil is positioned as a counter layer. A spacer is used to maintain a gap between the two layers, allowing for contact and separation when force or pressure is applied to the top layer.

\begin{figure}[h]
\centering
\includegraphics[width=0.8\linewidth]{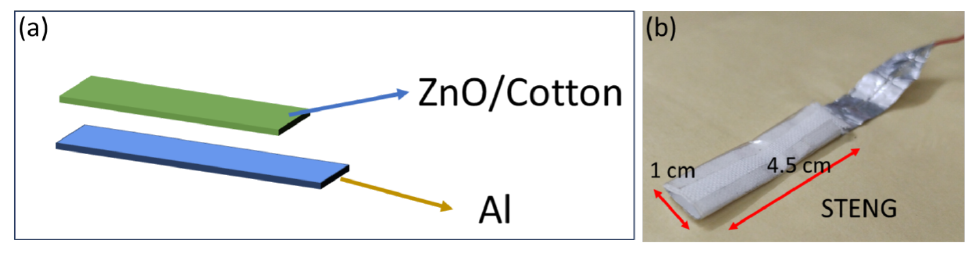}
\caption{Schematic representation and real image of the fabricated STENG or sensor}
\label{fig:Fabrication_of_STENG}
\end{figure}

\subsection{Signal acquisition}
To investigate the property of gesture recognition, a prototype of gloves attached with five sensors in each finger has been demonstrated using Arduino UNO. As shown in the Figure 2, five numbers from 1 to 5 have been recognized with the help of prototype gloves. In the recognition, it is observed that the generated output signals are different for each number. Also, by using the same prototype gloves, the study demonstrates the output signals for six alphabetic sign language from A to F.

\begin{figure}[h]
\centering
\includegraphics[width=0.8\linewidth]{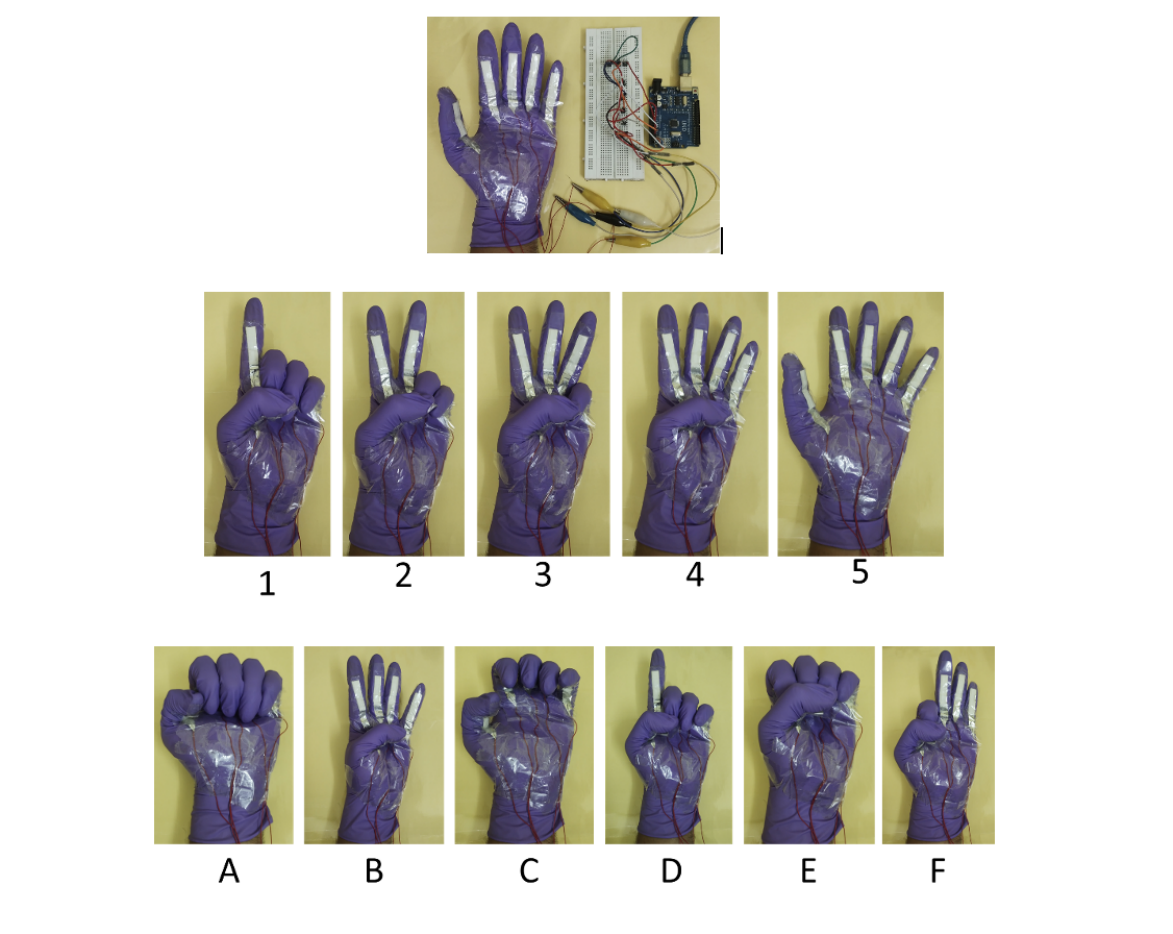}
\caption{Photograph of sensors attached to commercial nitrile gloves with variation of numeric and alphabetic sign languages}
\label{fig:sign}
\end{figure}

In the generated data sheet, the 1st, 2nd, 3rd, 4th, and 5th columns represent the generated output voltages for thumb, index, middle, ring and little finger, respectively. In Figure 3 a representative image has been shown for the generated output voltage due to the alphabetic sign language F.

\begin{figure}[h]
\centering
\includegraphics[width=0.8\linewidth]{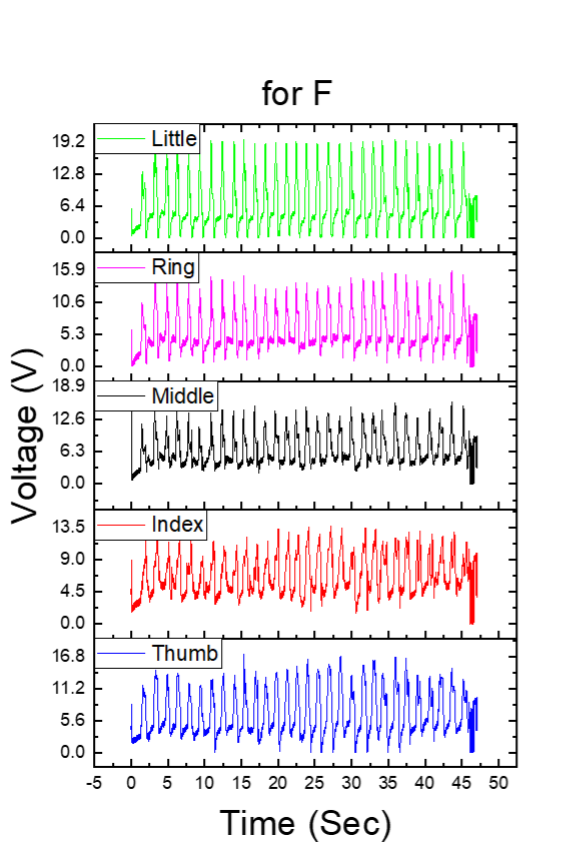}
\caption{Generated output voltage for sign language F}
\label{fig:sign_f}
\end{figure}

\subsection{Dataset}

The dataset comprises multivariate time-series data collected from five STENG sensors mounted on individual fingers of a glove. Each sensor continuously records analog readings proportional to finger flexion during sign language gesture execution. The dataset encompasses 11 sign classes: numerical digits 1 through 5 and alphabetical characters A through F.

Quality control procedures removed rows containing more than three zero values across sensors to eliminate sensor malfunction artifacts. Each sign class contains variable-length sequences ranging from 200 to 500 samples per gesture execution.

Data were collected from a single participant performing each gesture multiple times. The dataset was split into training (70\%), validation (15\%), and testing (15\%) subsets using random stratified splitting, ensuring no sequence appeared in more than one split.

\subsection{Data Preprocessing}

\subsubsection{Window-Based Sequence Segmentation}
Continuous signal data were segmented into fixed-length sequences without overlapping windows. Each file contains a total of $N$ data points and was split into $\lfloor N/w \rfloor$ samples with window size $w$. To retain the most current gesture data, the remaining data points that did not meet the minimum window requirement were removed from the start of each file. Different window configurations were evaluated. For example, a file with $N = 1028$ data points and window size $w = 50$ generates $\lfloor 1028/50 \rfloor = 20$ full sequences of 50 timesteps each (total 1000 data points), and the remaining 28 data points are discarded from the beginning.

\subsubsection{Sequence Normalization}
Two normalization strategies were used based on model architecture. Traditional ML methods used StandardScaler normalization (zero mean, unit variance) on flattened feature vectors in the training set~\cite{sklearn}. Deep learning techniques involved per-sequence normalization, where each sequence was independently normalized to zero mean and unit variance, to remove inter-subject baseline variability while preserving relative temporal patterns~\cite{batchnorm}.

\subsection{Feature Engineering}

\subsubsection{MFCC Feature Extraction}
Time-domain sensor inputs were converted to frequency-domain  using Mel-Frequency Cepstral Coefficients (MFCCs)~\cite{mfcc1,mfcc2}. The transformation was carried out in several steps:

\textit{Frame-based Processing:} Sensor signals were divided into overlapping frames with 32 samples per frame and hop length of 8 samples, resulting in 75\% overlap. Using Hamming windowing, $n_{frames} = 1 + (200-32)/8 = 22$ frames were generated for a 200-sample sequence.

\textit{Spectral Analysis:} Windowed frames were converted to the frequency domain by applying the Fast Fourier Transform (FFT), producing 17 frequency bins ($N_{FFT}/2+1$). The power spectrum was calculated as $P = |FFT|^2/N_{FFT}$, which shows the energy distribution across frequencies.

\textit{Mel-Scale Transformation:} A triangular Mel filter bank with $N_{\text{MELS}} = 40$ bands was constructed between 0 and 50~Hz (half of the sampling rate). 

The Mel scale conversion formulas are:
\begin{align}
M(f) &= 2595 \cdot \log_{10}\left(1 + \frac{f}{700}\right) \\
f(M) &= 700 \cdot \left(10^{M/2595} - 1\right)
\end{align}

The filter bank was applied through matrix multiplication:
\begin{equation}
\text{mel\_spectrum} = \mathbf{P} \times \mathbf{F}^T
\end{equation}
where $\mathbf{F}$ is the filter bank matrix $40 \times 17$.

\textit{Logarithmic Compression and DCT:} Log compression was calculated using $\log_{mel} = \log(mel_{spectrum} + \epsilon)$, where $\epsilon = 10^{-10}$ prevents numerical instability. The Type-II Discrete Cosine Transform (DCT) with orthonormal normalization was applied along the temporal axis, and the initial $N_{MFCC} = 12$ coefficients were retained.

To create a 3D tensor of shape (5 sensors, 22 frames, 12 coefficients) for multi-sensor processing, MFCC extraction was applied to each of the five sensors separately. Consequently, there are 1,320 MFCC features per 200-sample sequence.

\subsection{Data Augmentation}

Data augmentation was applied to both training and validation sequences to enhance model robustness, while the test set remained unaugmented. Four complementary techniques were implemented:

\textit{Gaussian Noise Injection:} Additive noise sampled from $\mathcal{N}(0, \sigma^2)$ where $\sigma = 0.02$, applied with 70\% probability to simulate sensor measurement uncertainty and environmental noise.

\textit{Time Warping:} Non-linear temporal deformation via cubic spline interpolation with $\sigma = 0.2$, applied with 60\% probability to simulate natural variations in gesture execution speed.

\textit{Magnitude Scaling:} Multiplicative scaling factor uniformly sampled from [0.9, 1.1], applied with 70\% probability to account for inter-subject variability in movement intensity.

\textit{Temporal Shifting:} Circular shift by random offset sampled from $[-10, +10]$ timesteps, applied with 50\% probability to address temporal misalignment in gesture start times.

Each training sample generated 2 additional augmented variants, yielding approximate $3\times$ dataset expansion. Augmentation was applied to raw time-domain sequences before MFCC extraction. Augmentation was employed for robustness testing, even though the final evaluation used the original test set.

\subsection{Model Architectures}

\subsubsection{Traditional Machine Learning}
Eight classical algorithms were evaluated on flattened, StandardScaler-normalized feature vectors: (1) Random Forest with 200 trees and max\_depth=20, (2) Gradient Boosting with 200 estimators and learning\_rate=0.1, (3) Support Vector Machine with RBF kernel and C=10, (4) K-Nearest Neighbors with k=5 and distance weighting, (5) Decision Tree with max\_depth=20, (6) Logistic Regression with L2 regularization and C=10, (7) Gaussian Naive Bayes, and (8) Linear Discriminant Analysis. Performance was assessed via 5-fold stratified cross-validation.

\subsubsection{Feedforward Neural Networks}
Four progressively deeper architectures were evaluated: SimpleNN $(128 \rightarrow 64 \rightarrow 32)$,  DeepNN $(256 \rightarrow 128 \rightarrow 64 \rightarrow 32)$,  WideNN $(512 \rightarrow 256 \rightarrow 128)$,  and UltraNN $(512 \rightarrow 256 \rightarrow 128 \rightarrow 64 \rightarrow 32)$ . All architectures used ReLU activations, batch normalization after each hidden layer, and dropout (0.2-0.5) for regularization. The models were trained with Adam optimizer (lr=0.001), batch size 64, maximum 200 epochs with early stopping (patience=20).

\subsubsection{LSTM-Based Temporal Models}
Two recurrent architectures were implemented: (1) Standard LSTM with stacked layers ($(128 \rightarrow 64 )$ units), batch normalization, dropout (0.3-0.4), and dense layers (128→64) before softmax output; (2) Attention-LSTM with bidirectional LSTM (128 units, $return\_sequences=True$), multi-head self-attention mechanism, global average pooling, and dense layers ($(128 \rightarrow 64 )$). Both models were trained with Adam optimizer (lr=0.001), batch size 32, maximum 200 epochs, and early stopping (patience=20).

\subsubsection{Multi-Sensor MFCC CNN-LSTM}
A parallel processing architecture handled the 5-sensor MFCC tensor. Five independent convolutional branches processed each sensor's MFCC sequence: Conv1D(12→64, kernel=3) → BatchNorm → ReLU → MaxPool1D(2) → Conv1D(64→128, kernel=3) → BatchNorm → ReLU → AdaptiveAvgPool1D(1). Branch outputs were concatenated (640 features) and fed through a fusion network: Dense(640→512) → LayerNorm → ReLU → Dropout(0.5) → Dense(512→256) → LayerNorm → ReLU → Dropout(0.4) → Dense(256→11). The model employed AdamW optimizer ($lr=0.001, weight\_decay=1\times10^{-4}$), Focal Loss ($\alpha=1$, $\gamma=2$), CosineAnnealingWarmRestarts scheduler ($T_0=10$, $T_{mult}=2$), gradient clipping ($max\_norm=1.0$), batch size 64, and maximum 150 epochs with early stopping (patience=30).

\section{Experimental Results and Discussion}
\label{sec:results}

\subsection{Overall Performance Comparison}

Table~\ref{tab:results} presents the test set performance of all evaluated approaches.

Table~\ref{tab:priorworks} elaborates the previously done work in this domain with different sensors.

\begin{table}[h]
\caption{Comparative Performance of All Approaches}\label{tab:results}
\centering
\begin{tabular}{|l|c|c|c|c|}
\hline
\textbf{Model} & \textbf{Accuracy} & \textbf{Precision} & \textbf{Recall} & \textbf{F1-Score} \\
\hline
MFCC CNN-LSTM & 93.33\% & 95.56\% & 93.33\% & 92.89\% \\
\hline
LSTM (Window 50) & 84.13\% & 83.40\% & 84.13\% & 82.62\% \\
\hline
Comprehensive LSTM & 82.54\% & 83.96\% & 82.54\% & 82.72\% \\
\hline
Random Forest & 70.38\% & 70.53\% & 70.38\% & 70.35\% \\
\hline
\end{tabular}
\end{table}

\begin{table}[h]
\caption{Comparison with Prior Glove-Based Sign Language Recognition Works}\label{tab:priorworks}
\centering
\renewcommand{\arraystretch}{1.2}
\begin{tabular}{|p{1.5cm}|p{1.4cm}|c|c|p{1.5cm}|}
\hline
\textbf{Reference} & \textbf{Sensor Type} & \textbf{Gestures} & \textbf{Best Accuracy} & \textbf{Method} \\
\hline
Ji et al. \cite{ji2023} & IMU (16-node) & 20 dynamic & 98.85\% & Attention-BiLSTM \\
\hline
Zhu et al. \cite{zhu2024} & Fiber optic & 12 & $\sim$95\% & ML-assisted \\
\hline
\textbf{This work} & \textbf{STENG (ZnO)} & \textbf{11 static} & \textbf{93.33\%} & \textbf{MFCC CNN-LSTM} \\
\hline
\end{tabular}
\end{table}

\begin{figure}[h]
\centering
\includegraphics[width=0.8\linewidth]{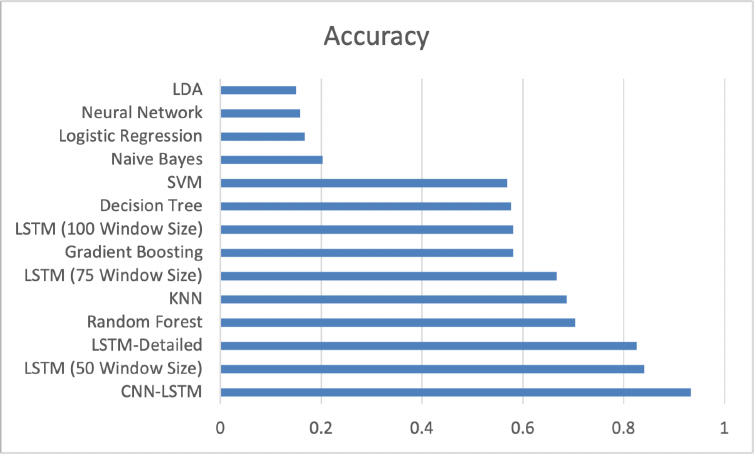}
\caption{Comparative performance of all models across different approaches}
\label{fig:all_model}
\end{figure}

Figure~\ref{fig:all_model} compares the accuracy of all trained models.

\subsection{MFCC-based CNN-LSTM Performance}

\begin{figure}[h]
\centering
\begin{subfigure}[t]{0.48\linewidth}
\centering
\includegraphics[width=\linewidth]{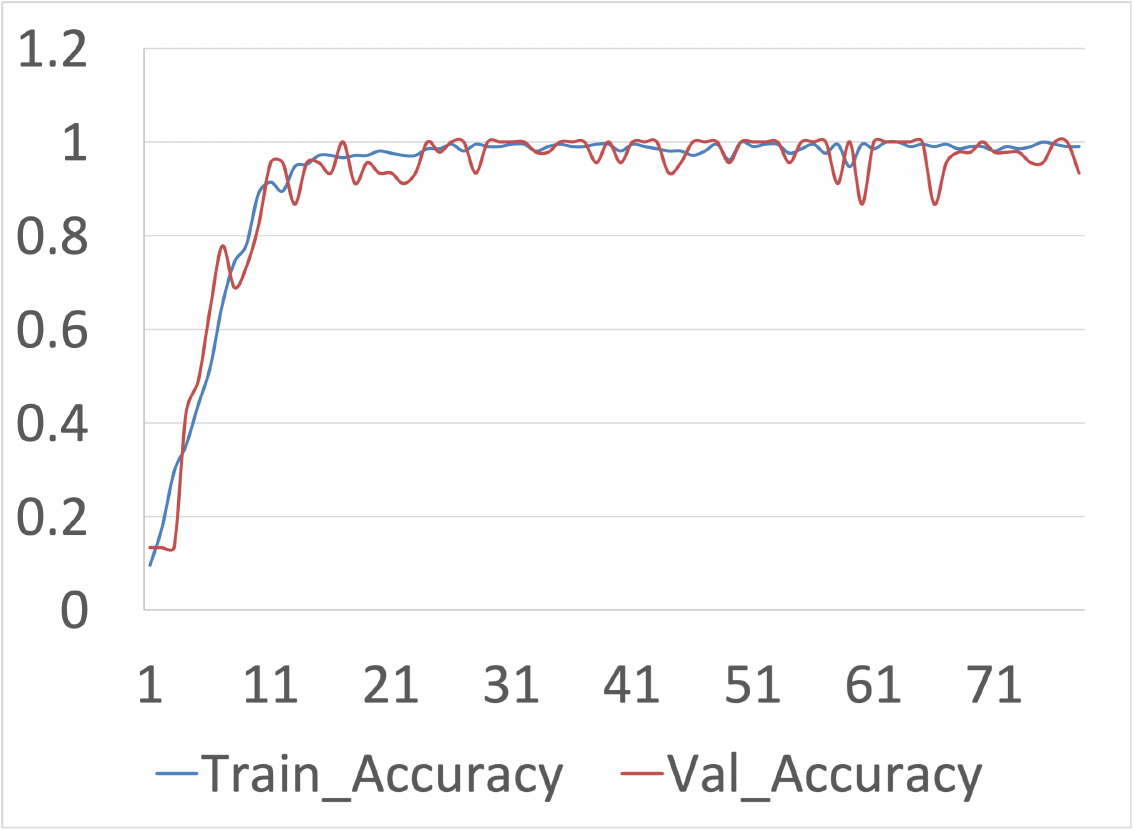}
\caption{Training and validation accuracy}
\label{fig:mfcc_accuracy}
\end{subfigure}
\hfill
\begin{subfigure}[t]{0.48\linewidth}
\centering
\includegraphics[width=\linewidth]{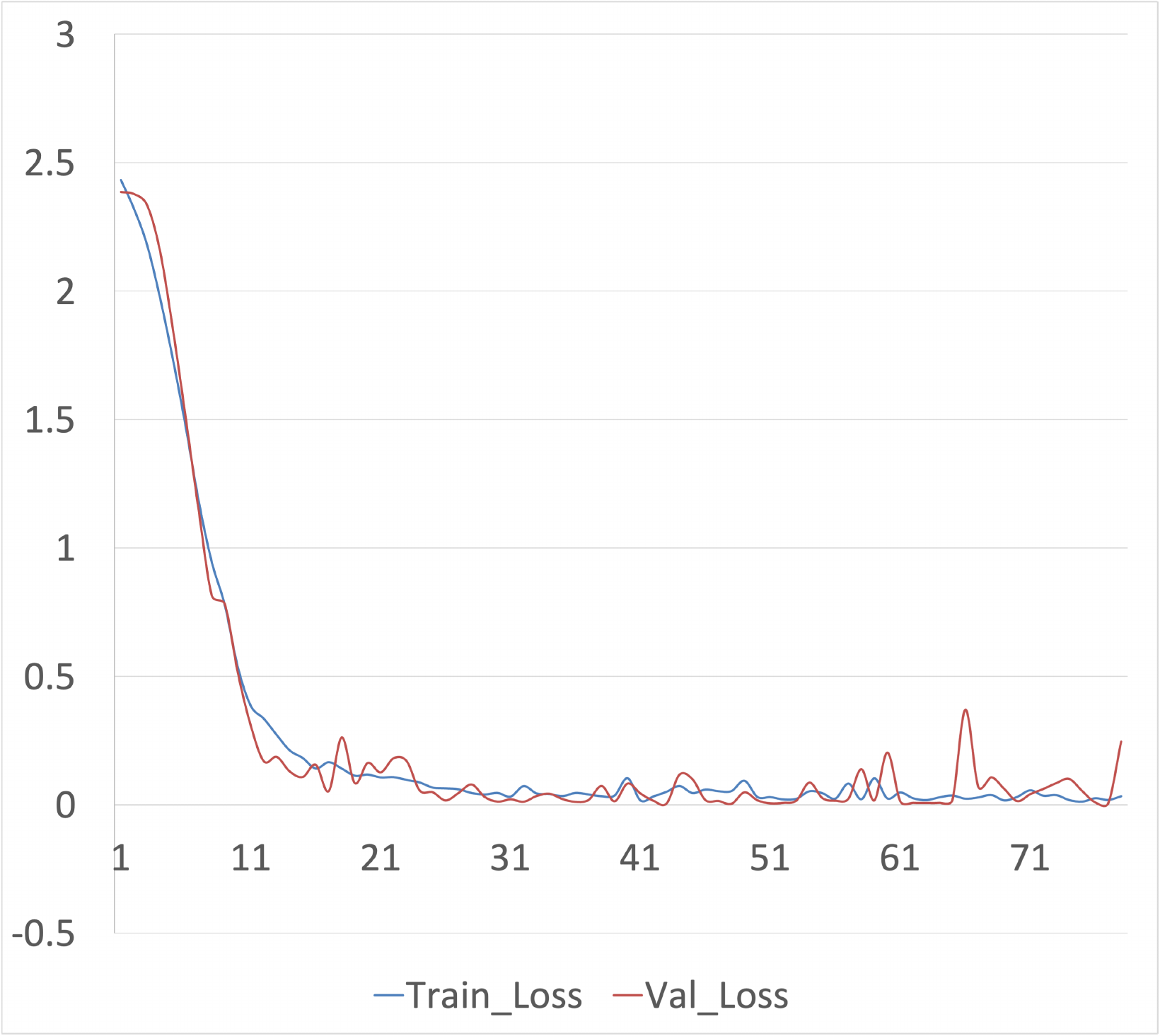}
\caption{Training and validation loss}
\label{fig:mfcc_loss}
\end{subfigure}
\caption{MFCC CNN-LSTM training performance across 78 epochs}
\label{fig:mfcc_training}
\end{figure}

The MFCC CNN-LSTM architecture achieved 93.33\% accuracy with exceptional precision (95.56\%), demonstrating a 23 percentage point improvement over the best traditional ML method. This superiority stems from two factors: (1) MFCC features capture frequency-domain patterns invariant to gesture execution speed variations, transforming temporal variations into spectral patterns, and (2) the parallel processing architecture processes each sensor's features separately through independent convolutional branches (32→64→128 channels), enabling sensor-specific frequency pattern learning before attention mechanisms aggregate information. The high precision-recall ratio (95.56\% vs. 93.33\%) indicates conservative classification behavior with minimal false positives a desirable characteristic for assistive technology applications where misclassification could lead to communication errors. The model exhibits bias toward high-confidence predictions, only classifying when strong evidence exists for a certain sign class. The consistent generalization gap between training (99.52\%) and testing (93.33\%) demonstrates effective regularization through batch normalization and dropout (0.3-0.4), minimizing overfitting despite the complex architecture. Figure~\ref{fig:mfcc_training} shows the performance of training and validation in 78 epochs.

\begin{figure}[h]
\centering
\includegraphics[width=0.9\linewidth]{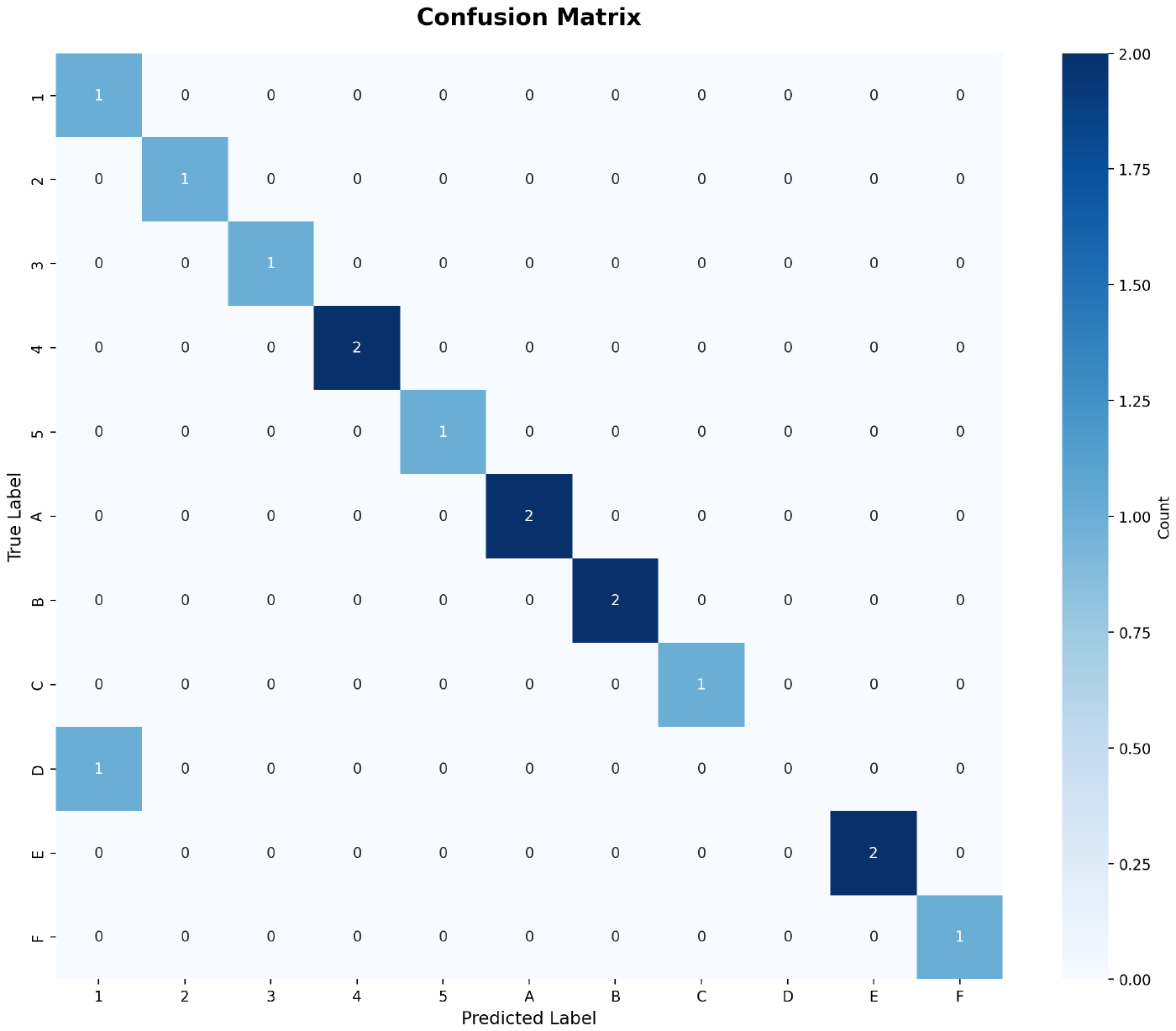}
\caption{Confusion matrix of MFCC CNN-LSTM on the test set}
\label{fig:confusion_matrix}
\end{figure}

Figure~\ref{fig:confusion_matrix} illustrates the confusion matrix of the MFCC CNN-LSTM model on the test set, showing each of the 11 gesture classes' per-class classification performance.

\subsection{Window Size Impact}

Table~\ref{tab:window} demonstrates the critical impact of window size on LSTM performance.

\begin{table}[h]
\caption{Performance Across Different Window Sizes}\label{tab:window}
\centering
\renewcommand{\arraystretch}{1.2}
\setlength{\tabcolsep}{3pt}
\begin{tabular}{|c|c|c|c|c|c|c|}
\hline
\textbf{Window} & \textbf{Accuracy} & \textbf{Precision} & \textbf{Recall} & \textbf{F1-Score} & \textbf{Epochs} & \textbf{Chunks} \\
\textbf{Size} & & & & & & \\
\hline
50 & 84.13\% & 83.40\% & 84.13\% & 82.62\% & 121 & 415 \\
\hline
75 & 66.67\% & 62.67\% & 66.67\% & 60.20\% & 61 & 274 \\
\hline
100 & 58.06\% & 54.95\% & 58.06\% & 51.47\% & 88 & 205 \\
\hline
\end{tabular}
\end{table}
\begin{figure}[h]
\centering
\begin{subfigure}[t]{0.48\linewidth}
\centering
\includegraphics[width=\linewidth]{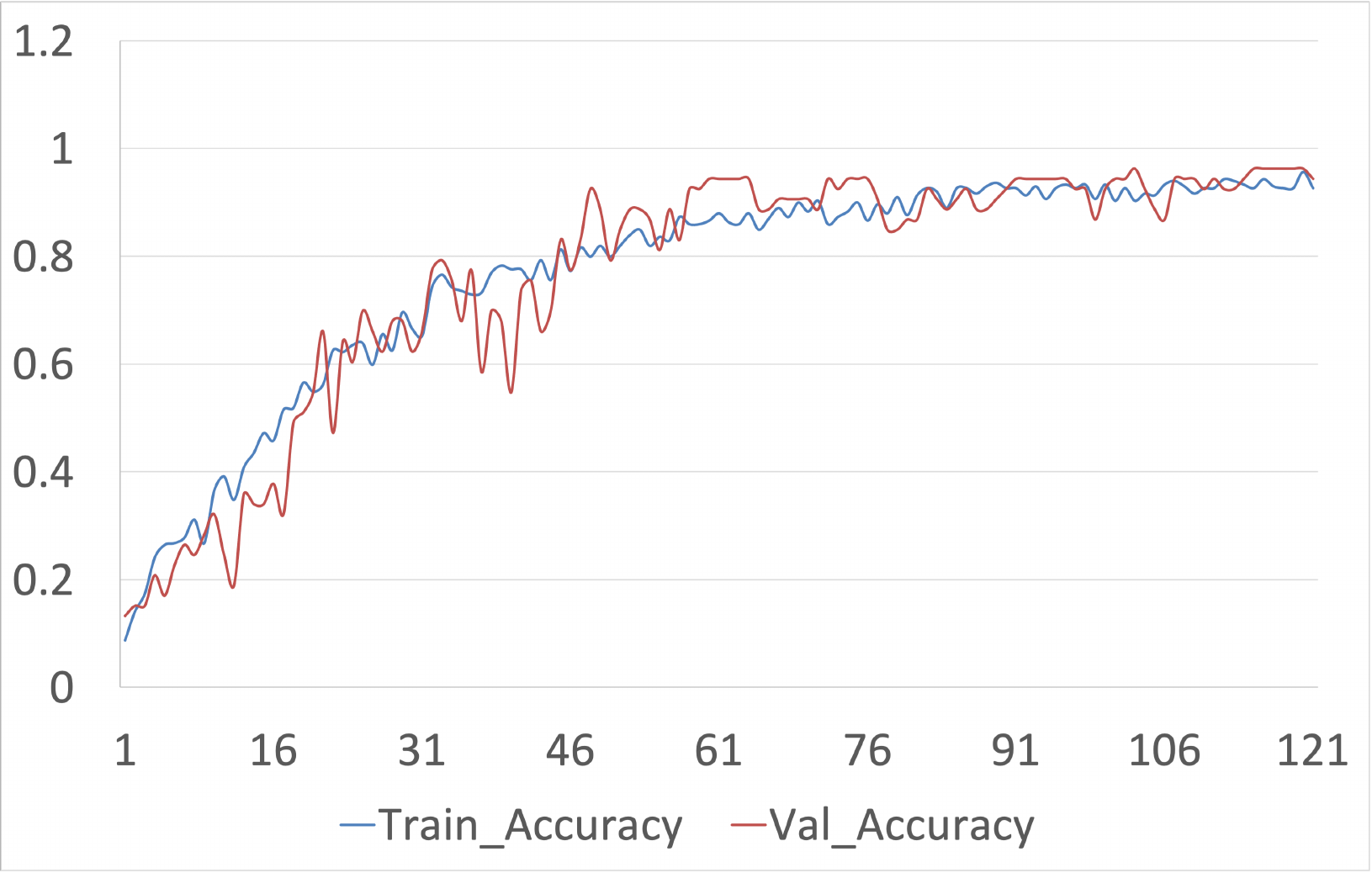}
\caption{Accuracy (window size 50)}
\label{fig:accuracy_50}
\end{subfigure}
\hfill
\begin{subfigure}[t]{0.48\linewidth}
\centering
\includegraphics[width=\linewidth]{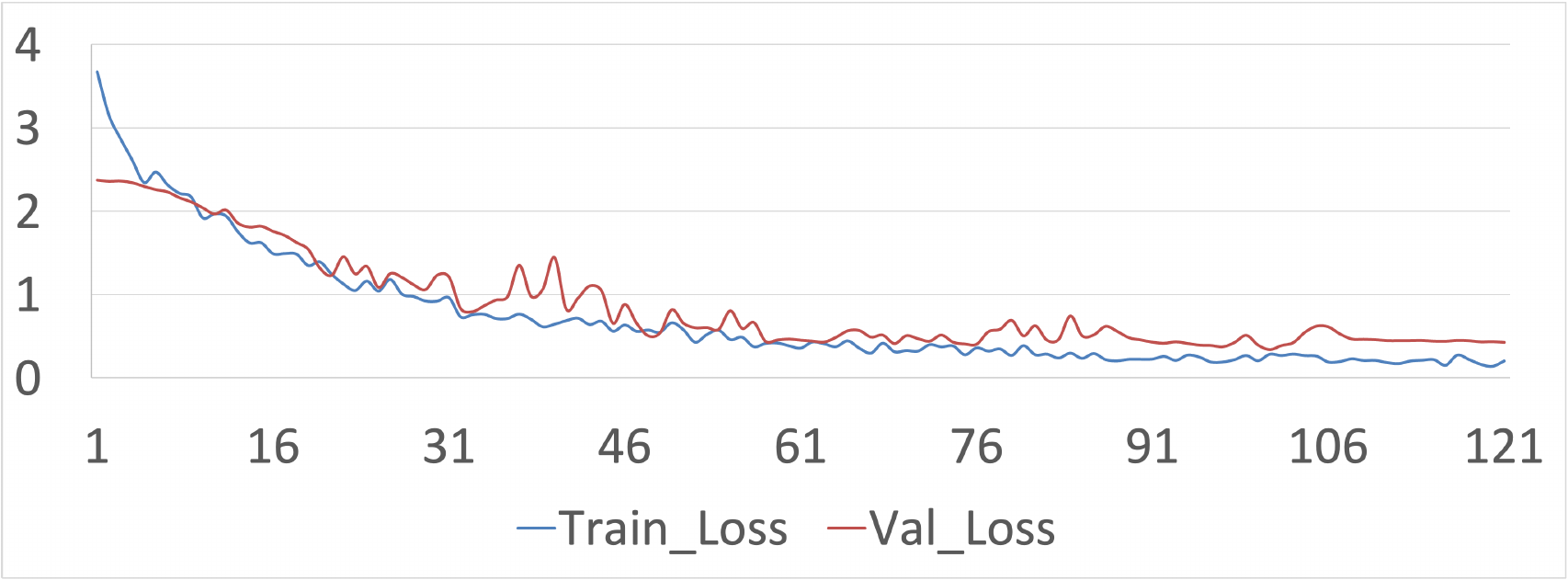}
\caption{Loss (window size 50)}
\label{fig:loss_50}
\end{subfigure}

\vspace{0.3cm}

\begin{subfigure}[t]{0.48\linewidth}
\centering
\includegraphics[width=\linewidth]{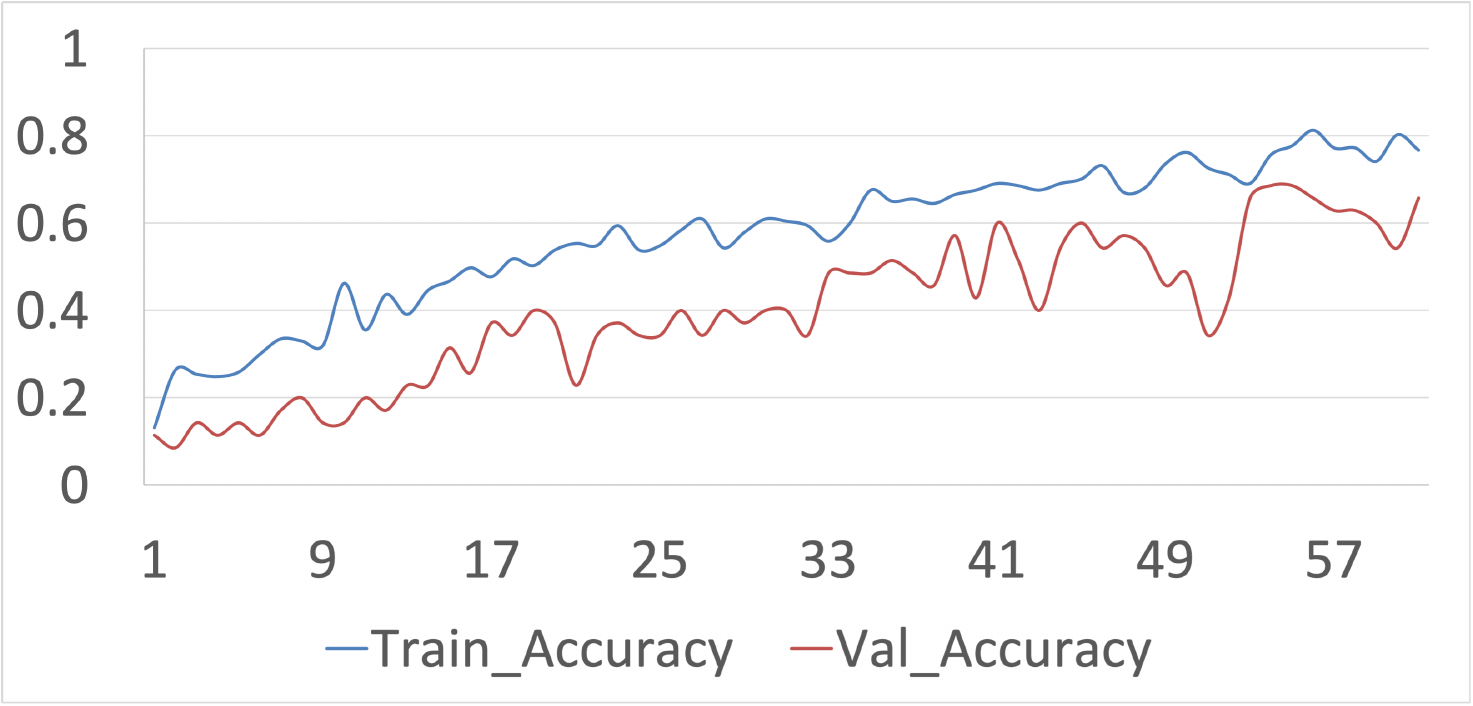}
\caption{Accuracy (window size 75)}
\label{fig:accuracy_75}
\end{subfigure}
\hfill
\begin{subfigure}[t]{0.48\linewidth}
\centering
\includegraphics[width=\linewidth]{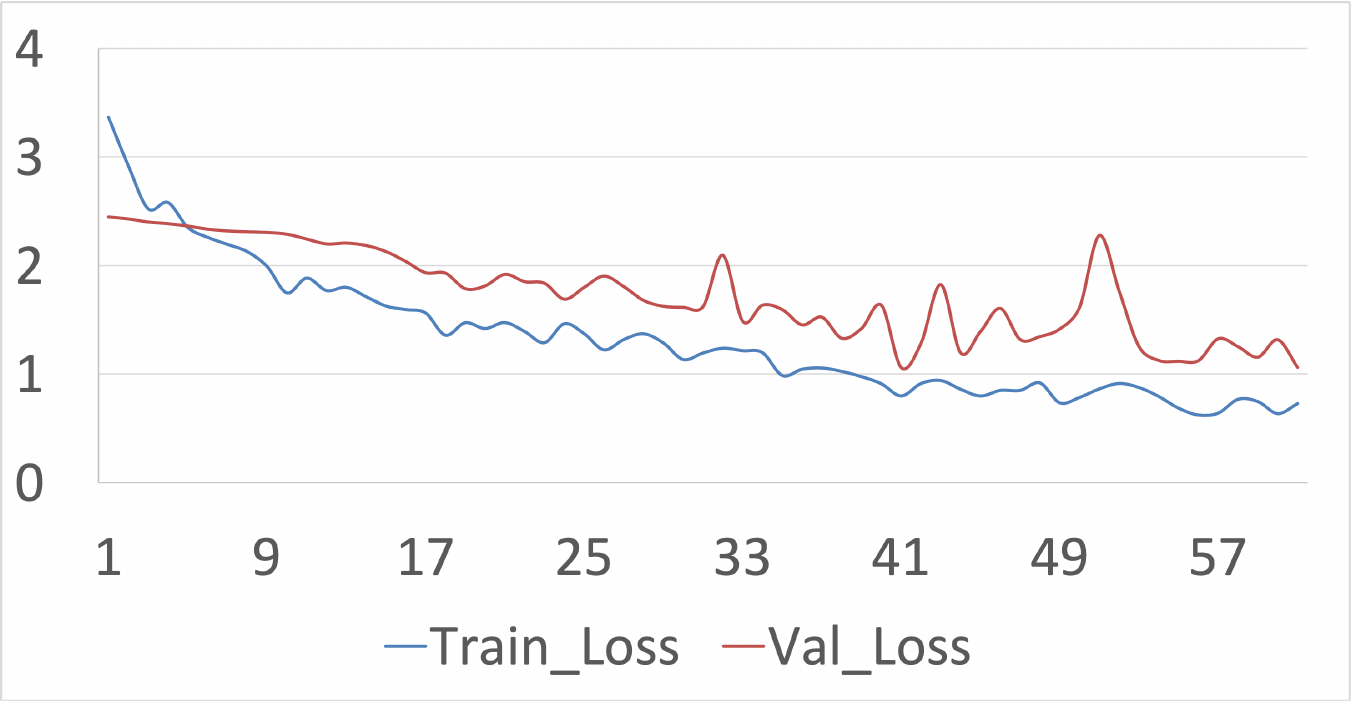}
\caption{Loss (window size 75)}
\label{fig:loss_75}
\end{subfigure}

\vspace{0.3cm}

\begin{subfigure}[t]{0.48\linewidth}
\centering
\includegraphics[width=\linewidth]{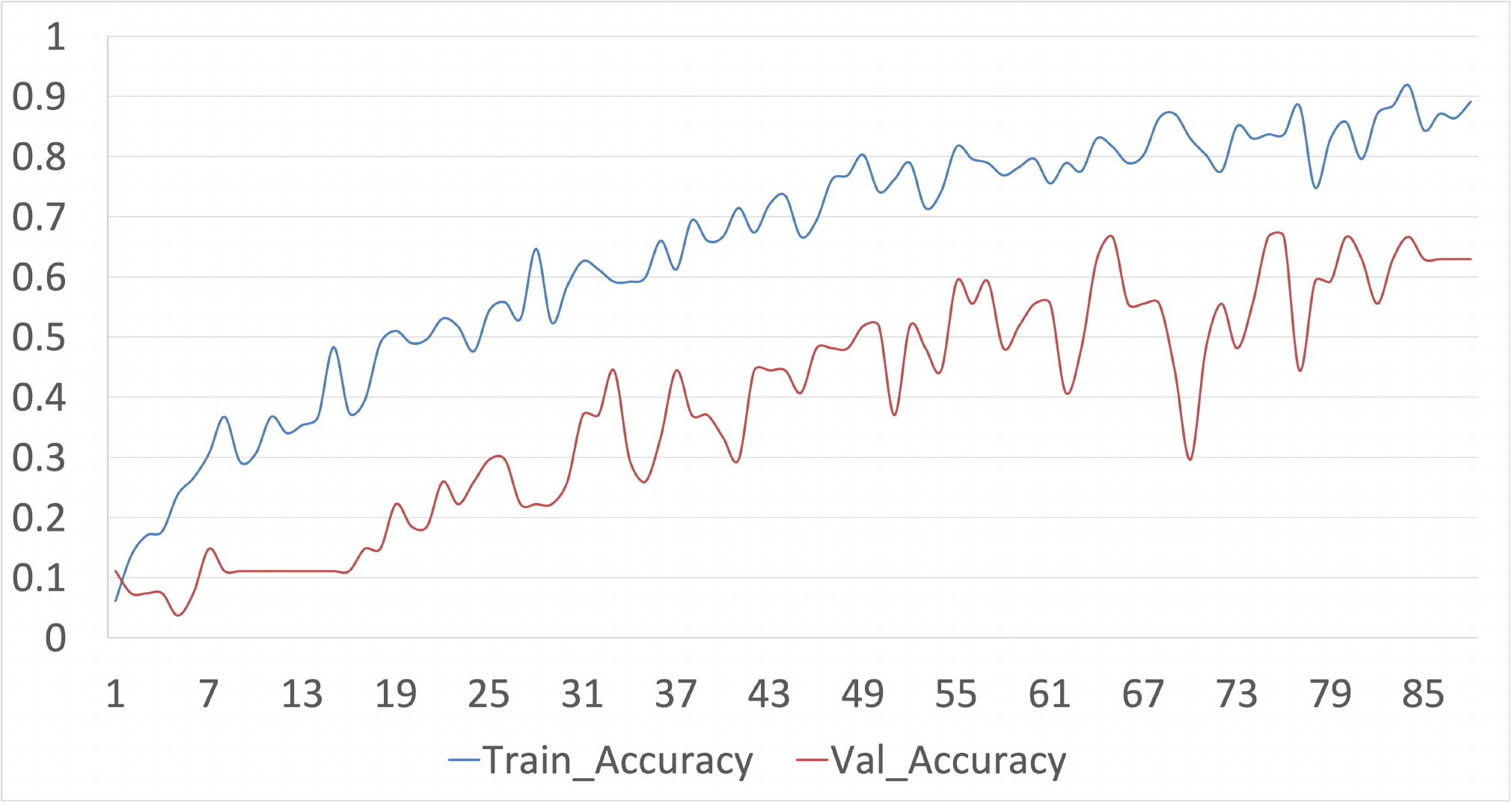}
\caption{Accuracy (window size 100)}
\label{fig:accuracy_100}
\end{subfigure}
\hfill
\begin{subfigure}[t]{0.48\linewidth}
\centering
\includegraphics[width=\linewidth]{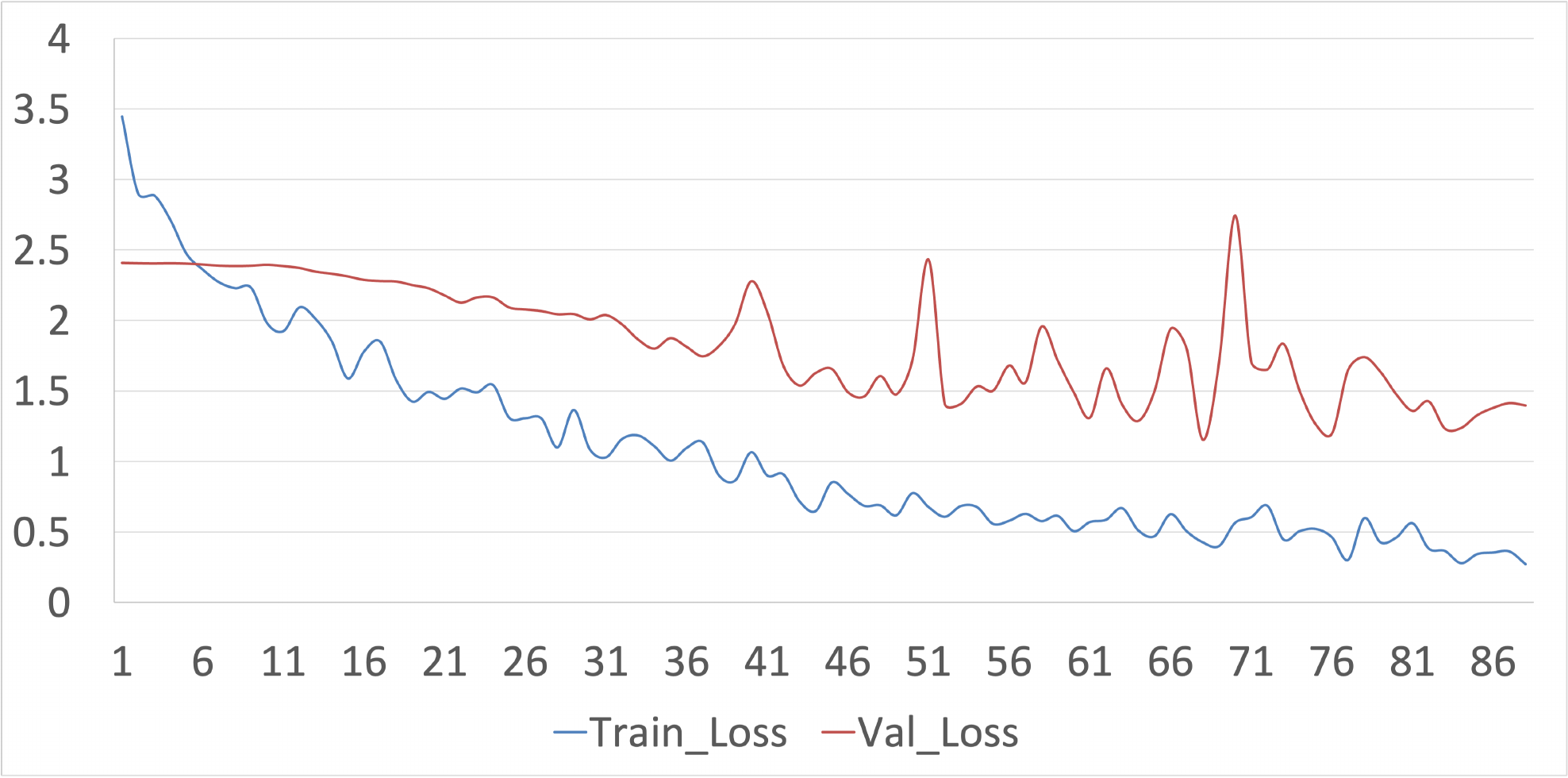}
\caption{Loss (window size 100)}
\label{fig:loss_100}
\end{subfigure}
\caption{Training accuracy and loss across epochs for different window sizes}
\label{fig:window_training_all}
\end{figure}

Figure~\ref{fig:window_training_all} illustrates the training dynamics across all three window configurations. The window size ablation study revealed two contributing factors to performance degradation with larger windows. First, the 100-timestep configuration produced only 205 chunks compared to 415 for the 50-timestep configuration—a 51\% reduction severely limiting model exposure to training examples. Increasing the chunk count improves temporal resolution rather than merely increasing the number of samples. However, additional windows may lead to more beneficial training instances. Second, longer sequences increase LSTM modeling difficulty, potentially exceeding the architecture's effective memory span and causing vanishing gradients. Training dynamics analysis revealed that the 50-timestep model showed stable validation performance throughout 121 epochs, while larger window sizes exhibited erratic convergence behavior, representing an optimal compromise between sufficient temporal context and adequate training data volume while aligning with typical sign gesture execution durations.

\subsection{Traditional Machine Learning Performance}

Table~\ref{tab:ml} presents the top-performing traditional ML models.

\begin{table}[h]
\caption{Traditional ML Performance (Top 5 Models)}\label{tab:ml}
\centering
\begin{tabular}{|l|c|c|c|c|}
\hline
\textbf{Model} & \textbf{Accuracy} & \textbf{Precision} & \textbf{Recall} & \textbf{F1-Score} \\
\hline
Random Forest & 70.38\% & 70.53\% & 70.38\% & 70.35\% \\
\hline
K-Nearest Neighbors & 68.64\% & 69.16\% & 68.64\% & 68.81\% \\
\hline
Gradient Boosting & 58.12\% & 58.61\% & 58.12\% & 57.99\% \\
\hline
Decision Tree & 57.60\% & 57.85\% & 57.60\% & 57.69\% \\
\hline
Support Vector Machine & 56.88\% & 60.10\% & 56.88\% & 57.20\% \\
\hline
\end{tabular}
\end{table}

\begin{figure}[h]
\centering
\includegraphics[width=0.7\linewidth]{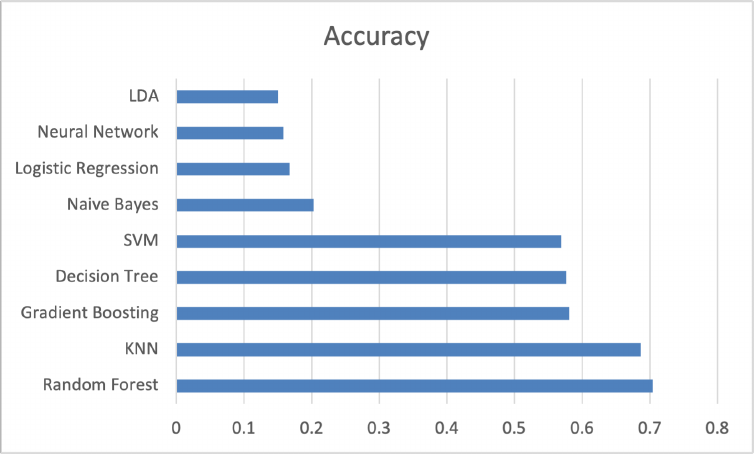}
\caption{Performance comparison of traditional ML models}
\label{fig:ml_accuracy}
\end{figure}

Figure~\ref{fig:ml_accuracy} provides a direct comparison of different ML model performance. Random Forest's ensemble decision-making over 200 trees achieved 70.38\% accuracy by handling high-dimensional feature spaces and capturing nonlinear decision boundaries. However, the 13.75 percentage point gap from the simplest LSTM approach demonstrates fundamental limitations for temporal sequence modeling. K-Nearest Neighbors (68.64\%, k=3) enabled instance-based classification but inability to capture long-range temporal connections restricted effectiveness compared to recurrent architectures. Decision Trees (57.60\%), Gradient Boosting (58.12\%), and Support Vector Machines (56.88\%) showed middling performance, struggling to model complex temporal relationships in high-dimensional time-series feature space. The simple feedforward neural network performed poorly (15.86\%), highlighting the necessity of recurrent connections for temporal modeling—a stark contrast with LSTM effectiveness.

\subsection{Comprehensive Neural Network Analysis}

\begin{figure}[h]
\centering
\begin{subfigure}[t]{0.48\linewidth}
\centering
\includegraphics[width=\linewidth]{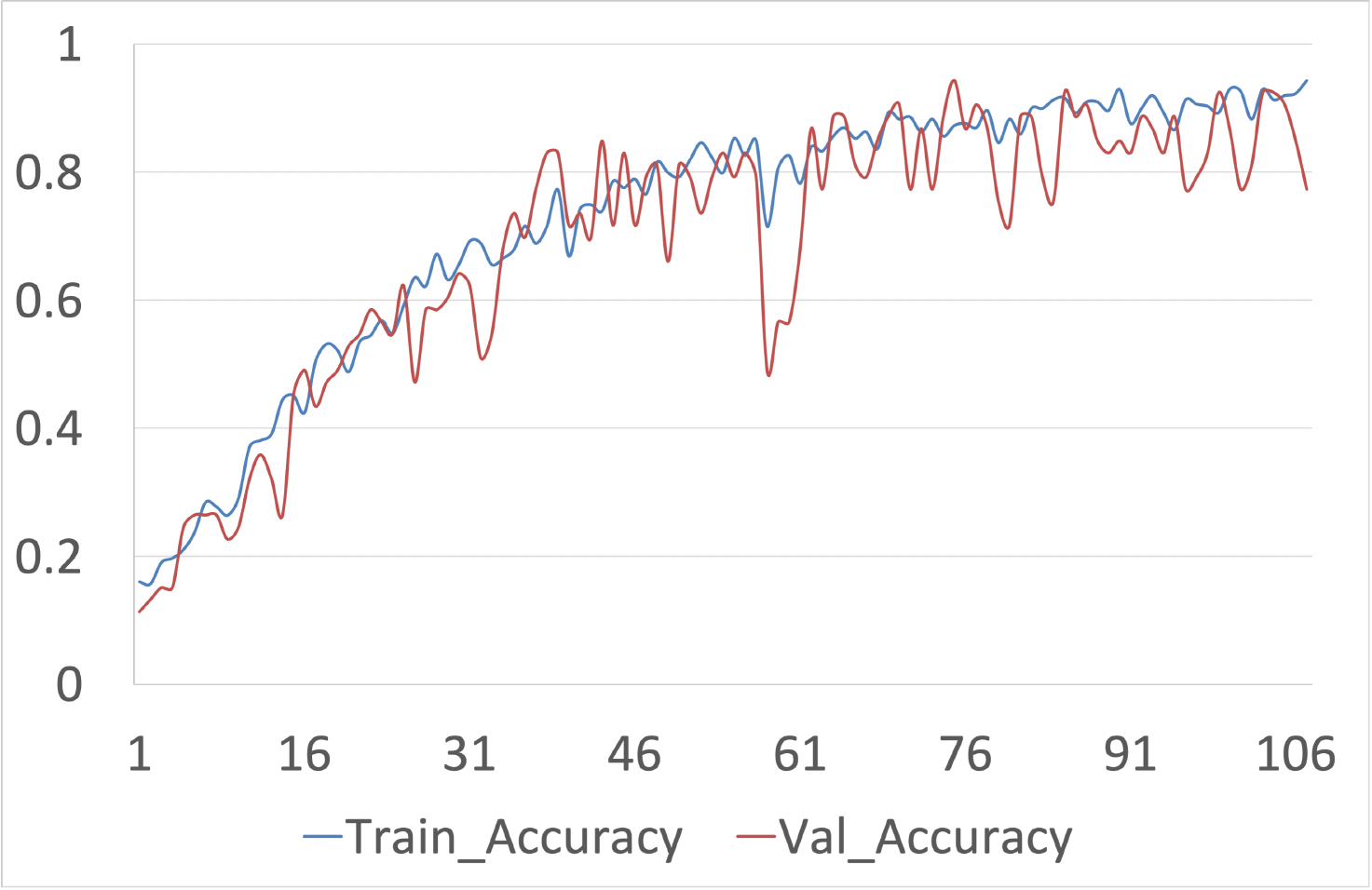}
\caption{Training and validation accuracy}
\label{fig:nn_accuracy}
\end{subfigure}
\hfill
\begin{subfigure}[t]{0.48\linewidth}
\centering
\includegraphics[width=\linewidth]{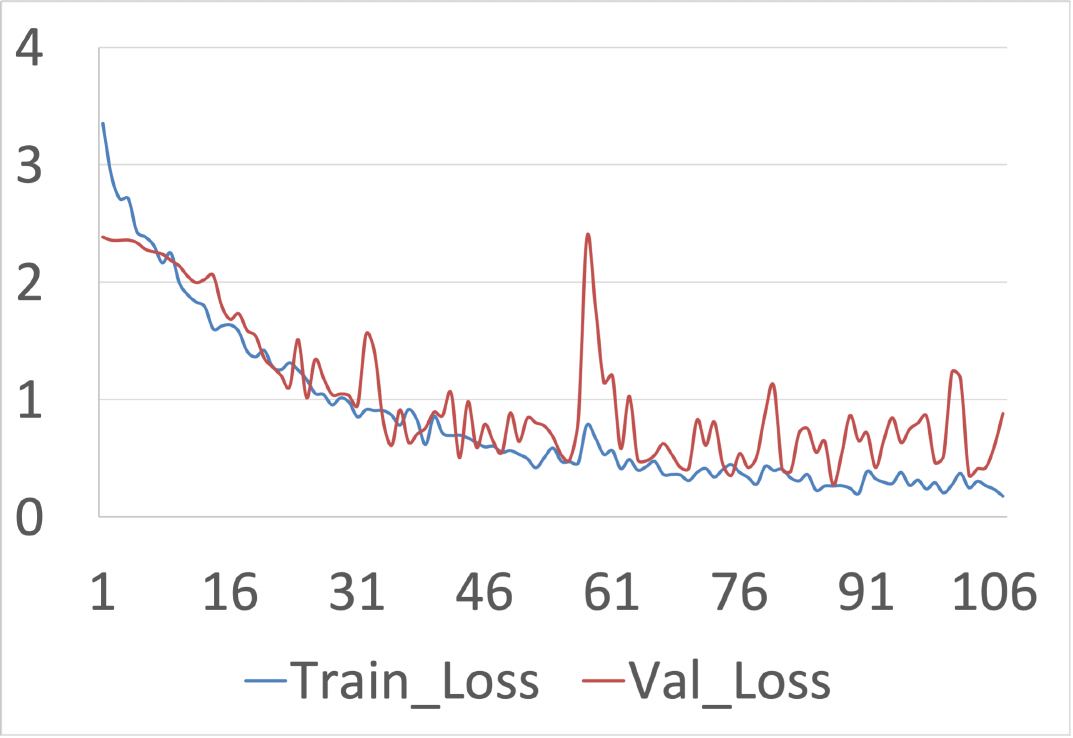}
\caption{Training and validation loss}
\label{fig:nn_loss}
\end{subfigure}
\caption{Comprehensive LSTM training performance across 107 epochs}
\label{fig:nn_training}
\end{figure}

Data augmentation strategies proved crucial for achieving high performance, employing four complementary techniques: time warping ($\sigma=0.2$), magnitude scaling (0.9-1.1 range), temporal shifting ($\pm$10 samples), and Gaussian noise injection ($\sigma=0.02$), yielding effective 3$\times$ dataset expansion applied to raw time-domain sequences before MFCC extraction. Figure~\ref{fig:nn_training} shows the comprehensive LSTM training performance across 107 epochs.

\section{Conclusion}
\label{sec:conclusion}

The primary limitation of this study is that all of the data was collected by a single individual using a single-made glove prototype. Since reported accuracy reflects within-subject performance under controlled conditions, it might not instantly apply to other users. Future studies will recruit a large number of participants and employ leave-one-subject-out cross-validation to fully assess inter-subject resilience. This paper elaborates on the in-depth ML and deep learning methods for data glove-based sign language recognition based on multivariate time-series sensor data.

The developed MFCC CNN-LSTM system achieves 93.33\% accuracy and 95.56\% precision, which opens a new era of sensor-based sign language recognition with 23 percentage points improvement over traditional ML methods. The findings of critical design ablation studies imply that an ideal balance between temporal context and training data availability exists in 50-timestep windows, speed-invariant representations generated by MFCC feature extraction, and comprehensive data augmentation strategies are the keys to successful generalization. The significance of domain-specific feature engineering for wearable sensors is demonstrated by the fact that frequency-domain processing outperforms time-domain methods. The capability of the parallel architecture to extract sensor-specific patterns before fusion is one indication of effective exploitation of the multi-sensor data structure. These results offer practical guidelines for developing robust assistive technology systems that can bridge communication barriers between the deaf and hearing communities. Future research will aim at expanding vocabulary to include complete sign language alphabets, exploring real-time implementation on embedded hardware platforms, and evaluating cross-subject generalization for practical deployment.

\end{document}